\documentclass[prl,aps,twocolumn,showpacs]{revtex4}
\usepackage{epsfig}
\usepackage{graphicx} 
\usepackage{amsmath}
\usepackage{amsfonts}
\usepackage{amssymb}

\newcommand{\ket}[1]{|#1\rangle}

\newcommand{\half}{\mbox{$\textstyle \frac{1}{2}$}}

\def\opone{\leavevmode\hbox{\small 1\kern -3.8pt\normalsize1}}

\begin{document}
\title{Measurement of geometric phase for mixed states using single
photon interferometry}
\author{Marie~\surname{Ericsson}, Daryl~\surname{Achilles}, Julio~T.~\surname{Barreiro},  David~\surname{Branning}$^*$, Nicholas~A.~\surname{Peters}, and Paul~G.~\surname{Kwiat}}
\affiliation{Department of Physics, University of Illinois at Urbana-Champaign, Urbana, IL 61801}
\date{January 14, 2004}
\begin{abstract}
Geometric phase may enable inherently fault-tolerant quantum
computation. However, due to potential decoherence effects, it is
important to understand how such phases arise for {\it mixed}
input states.  We report the first experiment to measure mixed-state 
geometric phases in optics, using a Mach-Zehnder interferometer, 
and polarization mixed states that are produced in two different ways: decohering pure
states with birefringent elements; and producing a
nonmaximally entangled state of two photons and tracing over one
of them, a form of remote state preparation. 
\end{abstract}
\pacs{03.65.Vf, 03.67.Lx, 42.65.Lm}

\maketitle When a pure quantum state undergoes a cyclic progression,
besides the dynamical phase which depends on 
the evolution Hamiltonian, it retains memory of its motion in
the form of a purely geometric phase factor~\cite{Panchar56,Ber84}. 
This pure-state geometric phase has
been experimentally demonstrated in various systems such as single
photon interferometry~\cite{Kwiat91}, two-photon 
interferometry~\cite{twophotonphase}, and NMR~\cite{Suter88}.
Recently, it has been proposed that fault-tolerant quantum
computation may be performed using geometric phases~\cite{geomet}, 
since they are independent of the
speed of the quantum gate and depend only on the area of the path
the state takes in Hilbert space. The next step is to investigate
the resilience of geometric phases to decoherence, and for this a
well defined notion of a mixed state geometric phase is needed.
Some properties of geometric phases for mixed states, proposed by
Sj\"oqvist et al.~\cite{sjoqvist}, have been recently investigated
in NMR~\cite{Du03}. Here, we report the first
experimental study of geometric phase for mixed quantum states
with single photons. Due to the exquisite control achievable
with optical qubits, we precisely map the behavior of the
phase for various amounts of mixture, yielding experimental data
in very good agreement with theoretical predictions.  
These results are particularly encouraging 
in light of recent work on scalable linear optics quantum
computation~\cite{klm}.

In order to measure a geometric phase, the dynamical phase has to be
eliminated. It can either be canceled, for example, using spin-echo technique
for spins in magnetic fields~\cite{Ekert00}, or one can parallel transport the
state vector in order to ensure that the dynamical phase is zero at all
times. The parallel transport condition for a particular state vector
$\ket{\Psi(t)}$ is $\langle \Psi(t)|\dot{\Psi}(t)\rangle=0$,
which implies that there is no change in phase when
$\ket{\Psi(t)}$ evolves to $\ket{\Psi(t+dt)}$, for some
infinitesimal change in time $t$. However, even though the state
does not acquire a phase locally, it can acquire a phase globally
after completing a cyclic evolution. This global phase is equal to
the geometric phase, and has its origin in the underlying
curvature of the state space. It is therefore resilient to certain
dynamical perturbations of the evolution, e.g., it is independent
of the speed (or acceleration) of evolution.

Uhlmann~\cite{uhlmann} was the first to describe mixed-state
geometric phases in a mathematical context where the parallel
transport of a mixed state is defined in a larger state space
which purifies the mixed state~\cite{purify}. In this approach the
number of parallel transport conditions for a
known $N\times N$ density matrix is $N^2$, but the time evolution
operator $U$ of such a density matrix has only $N$ free variables.
This approach can only be described in a
larger Hilbert space with the system and an attached ancilla
evolving together in a parallel manner~\cite{EPSBO2002}.

Sj\"{o}qvist et al. defined a mixed-state geometric phase where no
auxiliary subsystem is needed~\cite{sjoqvist,EPSBO2002}. This
phase can be investigated using an interferometer in which a
mixed state is parallel transported by a unitary operator in one arm; 
the output then interferes with the other arm, which
has no geometric phase. The parallel transport of a mixed state
$\rho=\sum_{k=1}^N p_k |k \rangle\langle k|$ is given by 
$\langle k(t)|\dot{k}(t)\rangle=0, \; \forall~k, \label{mixparalleltr}$
i.e., each eigenvector of the initial density matrix is parallel
transported by the unitary operator. The resulting $N$ conditions
uniquely determine the unitary operator and ensure the gauge
invariance of the geometric phase. One consequence of invariance
is that each eigenvector acquires a geometric phase $\gamma_k$,
and an associated interference visibility $v_k$. The total mixed-state 
geometric phase factor is then obtained as an average of the
individual phase factors, weighted by $p_k$:%
\begin{equation}%
v e^{i\gamma_g}  = \sum_k p_k v_k e^{i\gamma_k}. \label{gemphase}%
\end{equation}%

The polarization mixed state of a single photon can be represented
by its density operator, which can be written in terms of the
Bloch vector $\vec r$ and the Pauli matrices
$\vec\sigma=\{\sigma_x,\sigma_y, \sigma_z\}$, as
$\rho=\half(\opone+\vec r\cdot\vec\sigma)$. It represents a
mixture of its two eigenvectors with eigenvalues $\half (1\pm r)$.
The length of the Bloch vector $r$ gives the measure of the purity
of the state, from completely mixed ($r=0$) to
pure ($r=1$). For photons of purity $r$, Eq.~(\ref{gemphase}) becomes
\begin{equation}%
v e^{i\gamma_g} = \cos{(\Omega/2)} - i r\sin{(\Omega/2)}, \label{average}%
\end{equation}%
where $\Omega$ is the solid angle enclosed by the trajectory of one of the 
eigenvectors on the Bloch sphere with corresponding geometric phase 
$\Omega/2$ (the other eigenvector traverses the same 
trajectory, but in the opposite direction, leading to a geometric phase $-\Omega/2$).
From Eq.~(\ref{average}) we obtain the visibility and geometric phase, respectively,~\cite{sjoqvist}%
\begin{eqnarray}%
v&=&\sqrt{\cos^2{(\Omega/2)}+r^2 \sin^2{(\Omega/2)}}, {\rm and}\label{visibility} \\
\gamma_g&=&-\arctan \left( r\tan{(\Omega/2)} \right).\label{geomphase}
\end{eqnarray}%
Here $\gamma_g$ is measured in an interferometer by plotting the output intensity versus
an applied dynamical phase shift in one interferometer arm.  For pure states, the geometric phase
given by (\ref{geomphase}) reduces to half the solid angle ($\Omega/2$).

\begin{figure}%
\begin{center}%
\includegraphics[width=8.6cm]{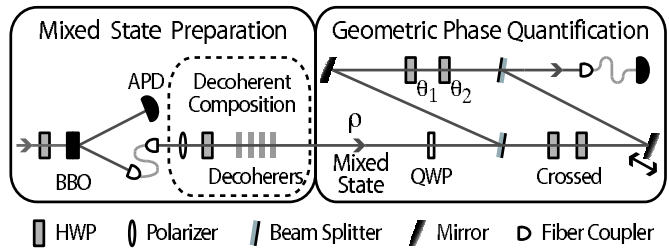}%
\end{center}%
\vspace{-0.63cm} \caption{Mixed state generation and interferometer to measure geometric phase. Mixed states are prepared via two methods: 1) tracing over the polarization of one photon
of a nonmaximally entangled polarization state and, 2) using an initial pure polarization state with
birefringent decoherers that couple polarization to photon arrival
time~(see dashed box)~\cite{footnote:laser}.  In the latter case, tracing over this time prepares a mixed
state.  Half-waveplates at $\theta_1$ and $\theta_2$ generate geometric phase but do not otherwise alter the transmitted polarization state.  Two crossed waveplates in the lower interferometer arm give no geometric phase, but compensate the optical path difference between the arms to achieve high visibility. The interferometer shape minimizes unwanted polarization changes
arising from non-normal mirror and beamsplitter reflections.} \label{fig:int}\vspace{-0.63cm}
\end{figure}

In our experiment, single-photon states are conditionally produced
by detecting one member of a photon pair produced in spontaneous
parametric downconversion (SPDC)~\cite{hong} (we also took data using coherent states from a diode laser).  Specifically, pairs of
photons at 670 nm and the conjugate wavelength 737 nm are
produced via SPDC by pumping Type-I phase matched BBO with an
Ar$^+$ laser at $\lambda=351$ nm.  By conditioning on detection of
a 737-nm ``trigger'' photon (with an avalanche photodiode after a
5-nm FWHM interference filter at 737 nm), the quantum state of the
conjugate mode is prepared into an excellent approximation of a
single-photon Fock state at 670 nm~\cite{Kwiat91,hong}, also with
wavelength spread $\delta\lambda\sim 5$~nm.  As shown in
Fig.~\ref{fig:int}, the 670-nm photons are coupled into a single-mode 
optical fiber to guarantee a single spatial mode input for
the subsequent interferometer.  A fiber polarization controller is
used to cancel any polarization transformations in the fiber.

The mixedness of the $670$-nm photons is set via two different
methods~\cite{KE}.  The first uses thick birefringent decoherers that couple
the single photon's polarization to its arrival time relative to the trigger
~\cite{berglund,footnote:laser}.  Consider a horizontally polarized ($| H \rangle$) and
a vertically polarized ($| V \rangle$) photon.  Assuming the decoherers delay
vertically polarized photons relative to horizontally polarized photons by more
than the photon's coherence length (given by $\lambda^2/\Delta\lambda\sim 90
\mu m$), upon detection of the trigger photon, $| H \rangle$ will in principle
be detected before $| V \rangle$.  Tracing over the timing information during
state detection erases coherence between these distinguishable states; 
this is equivalent to irreversible decoherence~\cite{KE}.   

To guarantee a pure fiducial state for this method of generating
mixed states, a horizontal polarizer is placed after the
polarization controller, followed by a half-waveplate (HWP), and
finally the decoherers (four pieces of quartz of $\sim$3 cm total
thickness).  By rotating the HWP, the state can be
prepared in an arbitrary superposition $\cos{\theta}\ket{H} +
\sin{\theta}\ket{V}$.  The light is then sent through the
decoherers, effectively erasing the off-diagonal terms in the
density matrix, resulting in purity $r=|\cos{2\theta}|$.  In our experiment, the eigenstates of the net geometric 
phase {\it operator} are circular polarizations; therefore, 
before entering the interferometer, the quantum
state is rotated with a quarter-waveplate (QWP) into a mixture of left
($|L\rangle\equiv (|H\rangle +i |V\rangle)/\sqrt{2}$) and right
($|R\rangle\equiv (|H\rangle -i |V\rangle)/\sqrt{2}$) circular
polarized light.

Our second method to produce mixed-polarization single-photon states, 
a version of remote state preparation~\cite{bennett01}, is to trace
over one of the photons of a pair initially in a nonmaximally entangled
polarization state.  This state is prepared using two thin BBO crystals
oriented such that pumping with polarization $\theta_p$ produces a variable
entanglement superposition state $\cos{\theta_p}|HH\rangle
+\sin{\theta_p}|VV\rangle$~\cite{Kwiat99}.  Here, the first position
polarization label corresponds to the trigger photon (at 737 nm) while the
second corresponds to its partner (at 670 nm).  A polarization-insensitive 
measurement of the trigger photon prepares the partner in
the polarization mixed state $\rho_{670 nm}~=~\cos^2{\theta_p}|H\rangle \langle
H|~+~\sin^2{\theta_p}|V\rangle \langle V|$, with $r=|\cos{2\theta_p}|$.  
$\rho_{670 nm}$ is~then transported over the
single-mode fiber (still with the polarization controller so the fiber does not
alter the state).  As before, a QWP is used to rotate the photon's polarization
state to a mixture of $|R\rangle$ and $|L\rangle$ before entering the
interferometer.

\begin{figure}\vspace{-0.5cm}
\begin{center}%
\epsfig{figure=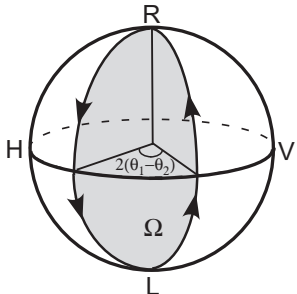,width=0.22\textwidth}
\end{center}\vspace{-0.9cm}
\caption{The solid angle $\Omega$ enclosed by the cyclic path of one 
eigenvector of the density matrix.  The other eigenvector traces the same path
but in the opposite direction, thus enclosing the solid angle $-\Omega$.  
$\Omega$ can be varied by adjusting $\theta_1-\theta_2$, the relative angle
between the optic axes of the two HWPs in the geometric phase arm of
the interferometer.} \label{fig:path}%
\vspace{-0.5cm}
\end{figure}

After any of the above mixed state preparations, the photon is
sent into a Mach-Zehnder interferometer (Fig.~\ref{fig:int}).  In
the upper arm, the Bloch vector $\vec r$ is evolved unitarily using
two half-waveplates with optic axes at $\theta_1$ and $\theta_2$,
respectively. The evolution can be illustrated
(Fig.~\ref{fig:path}) with one of the eigenvectors of the density
matrix, e.g., $|R \rangle $, traveling along two geodesics going
from $|R \rangle $ to $|L \rangle $ and back.  The trajectory
encloses a solid angle
$\Omega=4(\theta_1-\theta_2)$~\cite{footnote:wps}.  The other
eigenvector takes the same path but in the opposite direction, and
therefore encloses the solid angle $-\Omega$. For mixed states,
the length of $r$ is reduced, but the same solid angle is
subtended.  The resulting evolution fulfills the parallel
transport conditions for mixed states, and the induced 
geometric phase is obtained by substituting $\Omega/2=2\theta_1$
into equations (\ref{visibility}) and (\ref{geomphase}).  A
motorized rotation stage is used to set $\theta_1$ (to within
$0.01^{\circ}$) and thus, the geometric phase.

To measure $\gamma_g$ and $v$, we apply a dynamical phase shift
in the lower interferometer arm and measure the resulting
interference pattern both with a geometric phase (for several
settings of $\theta_1$~\cite{footnote:wedge}) and without
($\theta_1=0$).  The dynamical phase shift
is produced with a piezoelectric transducer (PZT) on the translation
stage on which the lower path mirror is mounted. By adjusting the
voltage across the PZT, the length difference ($\Delta L$) between
the arms is varied, giving the probability for the photon
to exit the interferometer to the detector as%
\begin{equation}%
P(\Delta L)=\frac{1}{2}\left( 1+\nu\cos\left(\frac{2 \pi
\Delta L}{\lambda}-\gamma_g\right) \right).%
\label{nonlinfit}
\end{equation}%
Photons are detected using an avalanche photodiode.  To conditionally prepare a
single-photon Fock state with the desired bandwidth, we count only coincident
detections (within a 4.5-ns timing window) with the trigger detector.  We
estimate the probability of two photons being present accidentally during a
given coincidence window is $3\times 10^{-6}$ for the decoherer method (using a 4-mm
thick BBO crystal) and $8\times 10^{-9}$ for the entanglement method (using two 0.6-mm
crystals).  Thus the ``accidental'' coincidence rate (e.g., between photons
corresponding to different pairs, or from detector dark counts) is
negligible, and has {\it not} been subtracted from the data.

\begin{figure*}
\begin{center}
\includegraphics[width=17.2cm]{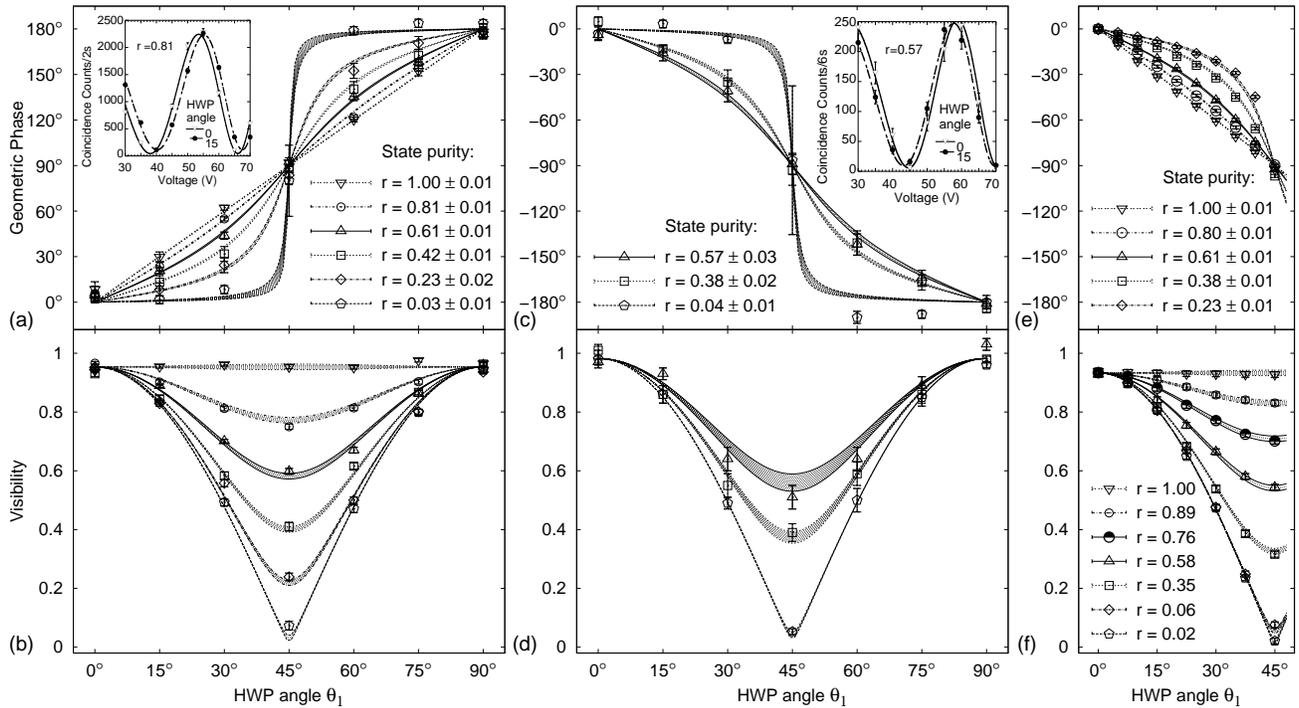}%
\vspace{-0.4cm}
\caption{The mixed state geometric phases and visibilities as a function of the
half-waveplate angle $\theta_1$. (a)-(b) The photons in the mixed
polarization state were produced with decohering quartz elements (see text).
(c)-(d) The mixed polarization state photons were produced by tracing
over one photon in a nonmaximally entangled state.  (e)-(f) The classical laser was decohered with an imbalanced polarizing interferometer.  The error bars are derived from the fit of the raw data to Eq.~\ref{nonlinfit}. The error in the 
theoretical curves shown results from
uncertainties in the determination of $r$, due to photon counting statistics (or intensity fluctuations for (e) and (f)).  The visibility theory curves are normalized
to the average visibility when $\theta_1=0$: 95$\%$ for (b), 98$\%$ for (d), and 93$\%$ for (e).  
The slightly imperfect visibility is largely due to imperfect interferometer
mode-matching.  Typical data is shown inset in (a) ($\theta_1=0^\circ$ and $\theta_1=15^\circ$ for $r=0.81$) and (c) ($\theta_1=0^\circ$ and $\theta_1=15^\circ$ and $r=0.57$).}
\label{fig:data}%
\end{center}
\vspace{-0.8cm}
\end{figure*}

Data is taken by varying the PZT voltage from $30$ to $70$ volts, 
in $5$ volt steps, giving slightly more than one period
of the interference pattern.  At each voltage, data is accumulated for $2$ s 
(decoherer method) or 6 s (traced-over entangled state method).  
We plot the number of coincidences as a function
of PZT voltage, and then fit a curve to extract the phase and
visibility information for each HWP
setting~\cite{footnote:volt}.  To calculate the phase difference
due to the geometric phase, we relate the data for each HWP setting
$\theta_1$ to the reference data with $\theta_1=0$ 
(see inset in Figs. 3(a) and 3(c))~\cite{footnote:averages}.

The experimental data are plotted in
Figs.~\ref{fig:data}(a)-\ref{fig:data}(f), along with theoretical
curves based on the measured purity of the photons.  To determine
the purity, we measure the $\ket{H}$ and $\ket{V}$ components of
the mixed state before the last quarter-waveplate in front of the
interferometer. Figs.~\ref{fig:data}(a)-(b) show the data for the
geometric phase and the visibility, respectively, for the
experiment where the single photons are decohered with thick
birefringent quartz. Figs.~\ref{fig:data}(c)-\ref{fig:data}(d) 
show the corresponding data when the mixture is due to
entanglement to the trigger photon.  Figs.~\ref{fig:data}(e)-\ref{fig:data}(f) show results from the coherent state, indicating that the data clearly fits the theoretical prediction and demonstrates that the single-photon geometric phase survives the correspondence principle classical limit~\cite{Kwiat91}.  The two geometric phase plots,
Fig.~\ref{fig:data}(a) and Fig.~\ref{fig:data}(c) are flipped
along the $x$ axis: in the first setup the input states possessed
larger right-circular polarization eigenvalues, while in the
second setup, left-circular polarization was dominant~\cite{footnote:flip}.

Fig. 3's error bars arise from the fitting program's uncertainty estimate of the phase and visibility from the raw fringes.  This error is consistent with the standard error obtained from repeating measurements 
four times to calculate the spread in the geometric phase and visibility.  We quantify how well the data fits the theory using a weighted reduced $\chi^2$-analysis.  
For the geometric phases (visibilities), we obtain average values of 0.98 (1.36) and 1.14 
(0.94) for the decoherer and entangled state 
preparations, respectively, indicating an excellent fit.  Also, the values of $r$
retrodicted from our fringe data agree with our direct measurements of $r$ within uncertainty.  

We report the first measurement of geometric phases for single
photons prepared in various polarization mixed states, created using two
different methods.  Specifically, we report a novel way of
creating decohered one-qubit states from entangled two-qubit
states, a simple version of remote state
preparation.  Both types of mixed states give
geometric phase and visibility data in very good agreement with
the theoretical predictions. Given the recent advances in linear
optical quantum computation~\cite{klm}, and continued interest in
geometric quantum computation~\cite{geomet}, our results indicate
that we have a good measure of the geometric phase for mixed
states, which in future work will enable the estimation of fault tolerance
in geometric quantum computation with linear optical elements.  
We also anticipate further experiments on non-unitarily evolved mixed
states~\cite{EPSBO2002} and non-Abelian geometric phases~\cite{wilczek84}, 
to ultimately realize a universal set of geometric gates for 
quantum computation~\cite{pachos}.

We thank J. B. Altepeter, E. R. Jeffrey, A. VanDevender,
and T.-C. Wei for valuable discussions and technical assistance.
M.E. acknowledges financial support from The Foundation
BLANCEFLORBoncompagni-Ludovisi, n$\acute{e}$e Bildt.  We recognize partial
support from the National Science Foundation (Grant \#EIA-0121568).


\begin{thebibliography}{99}
\bibitem{footnote:laser} The classical light is similarly mixed, using an unbalanced polarizing interferometer to separate $|H\rangle$ and $|V\rangle$ by more than the $\sim$1 m diode laser coherence length.
\bibitem{footnote:wps}  We can fix $\theta_2=0$, since $\Omega$ depends only on $\theta_1-\theta_2$.
\bibitem{footnote:wedge} The adjustable waveplate must not be wedged, or the path overlap will be ruined when the waveplate is rotated.
\bibitem{footnote:volt} Because the voltage applied to the PZT
creates a nonlinear increase in the path length, we fit our
data to a sinusoid with nonlinear dependence on the argument.  The PZT nonlinear fit parameters were determined by measuring dynamical phase fringes in our interferometer.
\bibitem{footnote:averages} To reduce the effect of dynamical phase drift, the reference data 
is taken both before and after each geometric phase-generating
HWP setting.  The data for each specific waveplate setting is used
to calculate the geometric phase relative to both the initial and final reference
data; the reported result is the average of the two calculations.
\bibitem{footnote:flip} The $r=0.04$ initial state was prepared 
more right circular polarized than left, opposite of the
initial states for the $r=0.38$ and $r=0.57$ trials.  Therefore,
for ease of comparison with these curves, the $r=0.04$ data curve 
is displayed flipped about the horizontal.
\end{thebibliography}
\end{document}